\documentclass{aa}
\usepackage{graphicx}
\usepackage{times}
\usepackage{mathptm}
\fontfamily{ptmcm}\selectfont

\begin{document}

\title{The radio halo in the merging cluster A3562}

\author{ T. Venturi         \inst{1}
	\and S. Bardelli    \inst{2}
	\and D. Dallacasa   \inst{1,3}
        \and G. Brunetti    \inst{1}
	\and S. Giacintucci \inst{2}
        \and R.W. Hunstead  \inst{4}
	\and R. Morganti    \inst{5}}
	
\offprints{Venturi T.}
\mail{tventuri@ira.cnr.it}

\institute{Istituto di Radioastronomia, CNR, Via\,Gobetti 101, I--40129, 
Bologna, Italy
\and Osservatorio Astronomico, Via Ranzani 1, I--40126 Bologna, Italy
\and Dipartimento di Astronomia, Universit\`a di Bologna, Via Ranzani 1,
I--40126, Bologna, Italy
\and School of Physics, University of Sydney, NSW2006, Australia 
\and NFRA, Postbus 2, A7990 Dwingeloo, The Netherlands}

\date{Received / Accepted }

\titlerunning{The radio halo in A3562}
\authorrunning{Venturi T. et al.}

\abstract{We present new VLA observations at 1.4 GHz confirming the
presence of a radio halo at the centre of the cluster A3562, in the core
of the Shapley Concentration. We also report a detailed 
multifrequency radio study of the head tail galaxy J1333--3141, which is
completely embedded in the halo emission. The radio halo has an irregular 
shape, and a largest linear size of $\sim$ 620 kpc,
which is among the smallest found in the literature.
The source has a steep spectrum, i.e. $\alpha_{843~MHz}^{1.4~GHz} \sim 2$, 
and its total radio power, P$_{1.4~GHz} \sim 2~\times10^{23}$ W Hz$^{-1}$,
is the lowest known to date.
The radio power of the halo and the X--ray parameters of the cluster,
such as L$_X$ and kT, nicely fit the correlations
found in the literature for the other halo clusters,
extending them to low radio powers.
We found that the total number of electrons injected in the cluster
environment by the head--tail source is enough to feed the halo, if we assume
that the galaxy has been radio active over a large fraction of its 
crossing time.
We discuss possible origins of the radio halo in the light of the 
two--phase model (Brunetti et al. 2001) and propose that the 
observed scenario is the result of a 
young source at the beginning of the reacceleration phase.
} 
\maketitle

\section{Radio halos and the case of A3562}

The existence of cluster radio halos and relics is well established,
and their number has considerably increased over the past few years
(e.g. Giovannini \& Feretti 2000, 2002; Govoni et al. 2001a; Bacchi et
al. 2003). At present about 40 objects are known,
supporting the evidence that many different components coexist
in clusters of galaxies over the whole cluster scale. In particular,
beyond the galaxies themselves and the hot thermal gas, visible in
the soft X--ray band, halos and relics reveal the existence of cluster
scale magnetic fields, with intensity of the order of $\sim~\mu$G,
and of relativistic electrons spread over the same large volume.

Both halos and relics are characterised by steep integrated radio 
spectra ($\alpha~>$ 1). In few well studied cases, halos
show a remarkable steepening of the spectrum going from
the centre to the peripheral regions. Halos are located
at the centre of the hosting cluster and are unpolarised, while
relics are usually found at the cluster outskirts, are more 
irregular in shape and show high fractional polarisation.
The linear size for both types of sources ranges from $\sim$ 0.5 up
to more than 1 Mpc (Feretti \& Giovannini 1996, Giovannini \& Feretti 2002).

The considerable number of radio halos known to date allows us to
search for correlations and other statistical properties. It is found that 
the radio power of halos strongly correlates with the cluster bolometric 
luminosity L$_X$, the intracluster gas temperature kT (Liang et al. 2000; 
Colafrancesco 1999; Feretti 2000) and the cluster mass M 
(Govoni et al. 2001a).
In particular, logP$_{1.4~GHz}$ increases with increasing
L$_X$, T and cluster mass, and the detection rate of radio halos 
is as high as $\sim~$ 30\% in clusters with L$_X~>~10^{45}$ erg s$^{-1}$
(Giovannini et al. 1999).  Finally, Govoni et al. (2001b)
found a spatial correlation between the halo radio brightness
and the X--ray brightness of the intracluster gas, suggesting 
some connection between these two components.
%
%

The major energy release ($\ge 10^{63} - 10^{64}$ erg s$^{-1}$) 
during cluster mergers has been invoked
to explain the formation of halos and relics. 
Absence of cooling flows, substructure and distortions in the galaxy 
and gas distribution, galaxy velocity structure, gas temperature 
gradients are all considered signatures of cluster mergers 
(see B\"ohringer \& Schuecker 2002 for a review on this topic). 
All these are known to be common properties of clusters
containing radio halos. 
Based on a quantitative analysis of the cluster dynamical state, 
Buote (2001) proposed that {\it massive} clusters experiencing 
{\it violent mergers}, i.e. those that have seriously disrupted the core,
are most likely to host a radio halo. On the other hand, many clusters with 
clear signs of merger do not possess a halo, suggesting
that our understanding of the whole picture is not yet complete.
In situ acceleration of relativistic electrons during a merger event,
may be due to merger shocks (e.g. Sarazin 1999) or to turbulence 
(e.g. Schlickeiser et al. 1987; Brunetti et al. 2001; Ohno et al.
2002). Recently, Gabici \& Blasi (2002) claimed that shock acceleration
provides efficient particle acceleration only in the case of minor
mergers. 

\medskip
A3562 (z=0.048, B--M type I and richness class 2)
is located in the central region of the Shapley Concentration, 
where major cluster
merging processes and group accretion are known to take place
(Ettori et al. 1997; Bardelli et al. 1996 and 1998). The cluster has
a temperature of kT = 5.1$\pm$0.2 keV, a bolometric luminosity 
L$_X~=~4.3\times10^{44}$ erg s$^{-1}$ and a total mass 
M$_{tot}~=~5.5\times10^{14}$ M$_{\odot}$ (Ettori et al. 2000).
The core region is characterised by a small cooling flow (Peres et al. 1998).
Beyond the central cooling flow, excess 
X--ray emission over the fitted $\beta$--model (significant at the 
95\% confidence level) was
detected with Beppo--SAX
west of the cluster centre, in the direction of A3558,
and at the extreme eastern periphery of the cluster, while
no significant excess hard X--ray emission over the thermal Bremsstrahlung 
component was found (Ettori et al. 2000).

Radio observations at 1.38 GHz and 2.36 GHz
(22 cm 13 cm respectively), carried out with the Australia Telescope
Compact Array, revealed the presence of 
a head--tail radio galaxy, J1333--3141, located at a projected distance
of $\sim~1^{\prime}$ from the cluster centre. Furthermore, hints of cluster 
scale emission were found, on the basis of a MOST 843 MHz image and 
inspection of the NVSS (Venturi et al. 2000, hereinafter V2000).
The radio survey revealed the presence of six more
radio galaxies located east of the cluster centre, in the
same region where the excess X--ray emission was noted.

\medskip
In this paper we confirm the presence of a radio halo in the 
cluster A3562, thanks to new 1.4 GHz Very Large Array (VLA)
observations, and perform an analysis of its properties
and origin. Furthermore, we carry out a thorough study of the 
head--tail radio galaxy J1333--3141, embedded in the halo emission, 
based on the data published in V2000 and on new VLA observations over
a wide range of frequencies.

The observations and data analysis are reported in Section 2 and 3; 
the discussion on the halo origin in the light of the two--phase
model (Brunetti et al. 2001), its connection with the
head--tail galaxy J1333--3141 and with the cluster merging in
A3562 are given in Section 4; conclusions are summarised in  
Section 5.

We will assume S $\propto~\nu^{-\alpha}$,  H$_o$ = 50 km s$^{-1}$Mpc$^{-1}$, 
q$_o$=0.5. With our cosmology, at the distance of 
A3562 1$^{\prime\prime}$ corresponds to 1.28 kpc.

\section{The head--tail radio galaxy J1333--3141}

The radio emission of the halo is blended with that of the
head--tail galaxy J1333--3141, both in the 1.4 GHz image presented
here and in the 843 MHz image (V2000). 
For this reason, in order to estimate the integrated spectral index 
of the halo between 843 MHZ and 1.4 GHZ and to study its properties, 
a detailed knowledge of the head--tail radio galaxy is also necessary.


J1333--3141 is associated with the cluster galaxy MT\,4108 
(v=14438 km/s and b$_J$ = 17.25), which is located at a projected
distance of $\sim~1^{\prime}$ from the central cD galaxy (V2000).

%
\begin{table*}[ht]
\caption{Log of the observations}
\label{log}
\begin{tabular}{lllcccc}
\hline
 &&&&&&\\
 Source \& Field & RA$_{J2000}$ & DEC$_{J2000}$ & u--v range & $\nu$ & 
$\Delta\nu$ & Duration \\
\\
 & h,m,s & $^{\circ}$,$^{\prime}$,$^{\prime\prime}$ & k$\lambda$ & GHz & 
MHz & hr \\
 &&&&&&\\
\hline
 &&&&&\\
A3562 \# 1 & 13 33 32.0 & $-31$ 41 00.0 & 0.15 -- 6     & 1.40 & 50 & 1 \\
J1333--3141 & 13 33 31.7 & $-$31 41 10.0 & 0.2 -- 15  & 0.33 & 6.25 & 0.2 \\
J1333--3141 & 13 33 31.7 & $-$31 41 10.0 &  3 -- 250  & 4.86 & 100  & 0.5 \\
J1333--3141 & 13 33 31.7 & $-$31 41 10.0 &  4 -- 400  & 8.46 & 100  & 0.5 \\

 &&&&&&\\
\hline
 &&&&&&\\
\end{tabular}
\end{table*}

\subsection{Radio Observations}
 
The source was observed at 330 MHz, 4.86 and 8.46 GHz 
with the VLA in the BnA configuration on 22 May 2002. 
Details of the observations are given in Table 1.

Standard calibration, imaging and data analysis were carried out 
with the NRAO AIPS package.

\begin{figure} [!ht]
\resizebox{\hsize}{!}{\includegraphics{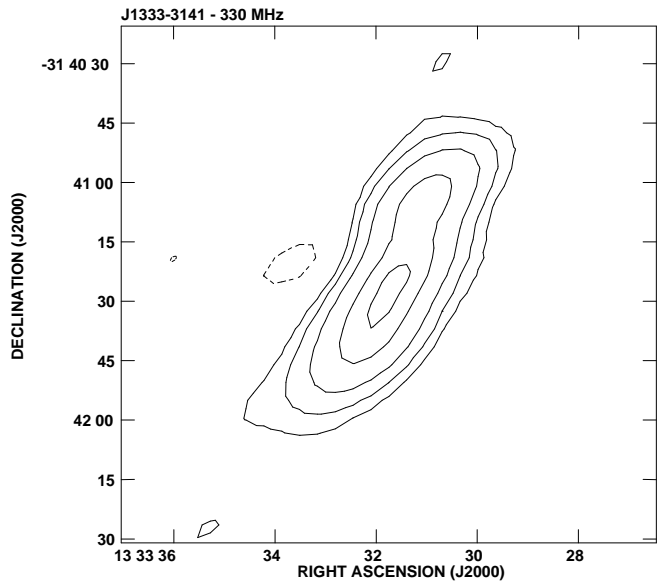}}
\caption{ VLA 330 MHz image of J1333--3141. The FWHM is 
$27.0^{\prime\prime} \times 12.5^{\prime\prime}$, p.a. $-52^{\circ}$.
Contours are: --8,8, 16, 32, 64, 90 mJy/beam. The peak
in the image is 99 mJy/beam.}
\label{j1333p}
\end{figure}

\begin{figure} [!ht]
\resizebox{\hsize}{!}{\includegraphics{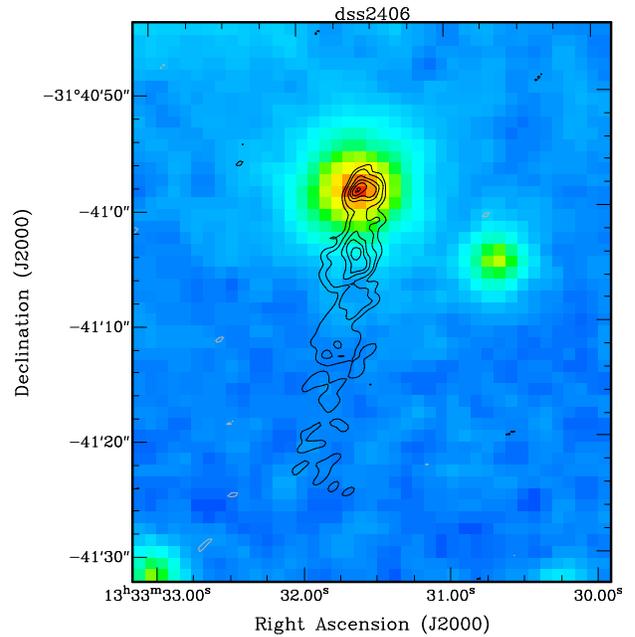}}
\caption{ VLA 4.86 GHz  image of J1333--3141 overlaid on the
optical DSS--2 red image. The FWHM is 
$1.46^{\prime\prime} \times 0.84^{\prime\prime}$, p.a. $-57^{\circ}$.
Contours are: --0.15, 0.15, 0.3, 0.6, 0.8, 1, 1.2 mJy/beam. The peak
in the image is 1.25 mJy/beam.}
\label{j1333c}
\end{figure}

\begin{figure} [!ht]
\resizebox{\hsize}{!}{\includegraphics{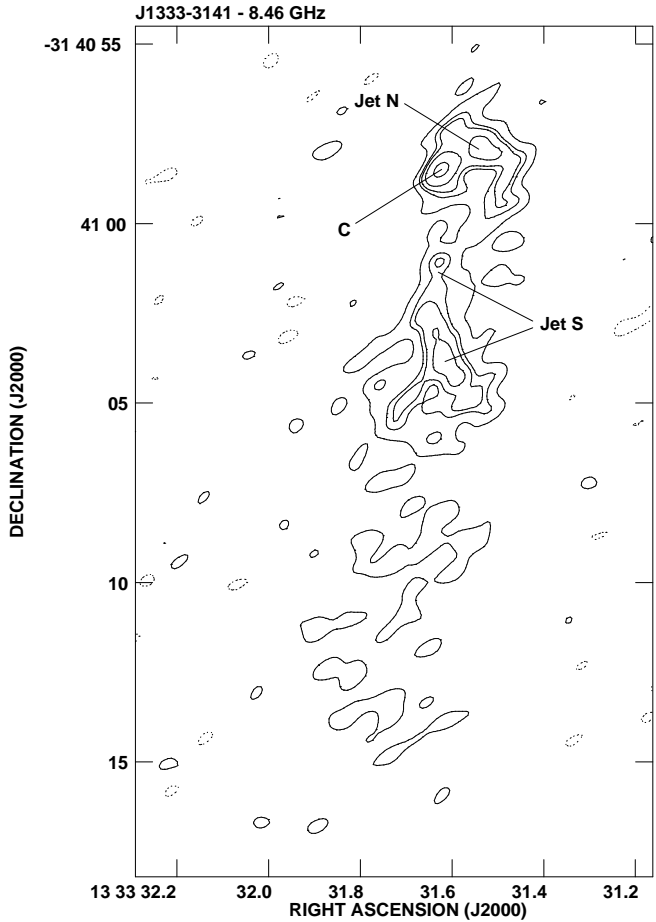}}
\caption{ 8.46 GHz image of J1333--3141. The FWHM is 
$0.88^{\prime\prime} \times 0.51^{\prime\prime}$, p.a. $-53^{\circ}$.
Contours are: --0.09, 0.09, 0.18, 0.23, 0.32, 0.54 mJy/beam. The peak
in the image is 0.64 mJy/beam.}
\label{j1333x}
\end{figure}

The 330 MHz image and the
final full resolution images at 4.86 GHz (overlaid on the DSS--2 red 
plate) and at 8.46 GHz are given respectively in Figs. \ref{j1333p}, 
\ref{j1333c} and \ref{j1333x}.
Details of all images are given in Table 2. 
The error on the total source flux density is $\sim$3\%
at 4.86 GHz and 8.46 GHz, based on gain variations on different scans
on the secondary calibrators. This error can be as high as 10\% at 330 MHz, 
due to the shorter duration of the observations and to rapid phase 
fluctuations introduced by the ionosphere at this low frequency.

A set of images obtained with natural weighting (not shown here) were also 
produced at 4.86 GHz and 8.46 GHz, in order to derive the 
integrated spectrum of J1333--3141 (see next subsection). We note 
that the total flux density in these natural weighted images is 
$\sim 5 - 10$\% higher than the values given in Table 2. 

Finally, for an accurate point--to--point spectrum evaluation, 
necessary for the study carried out in Sections 2.3 and 4, we used
the data presented here and re--analysed the 
ATCA data published in V2000 to produce various sets of images
using the same common portion of the u--v coverage, gridding and 
restoring beam $\theta_R$, as follows (see also Table 3 for a summary):

\noindent 
{\it (a)} images at 330 MHz, 1.38 GHz and 2.36 GHz, obtained using the data
points in the u--v range 0.5 -- 15 k$\lambda$, 
$\theta_R = 25^{\prime\prime} \times 12^{\prime\prime}$, p.a. $-50^{\circ}$;

\noindent
{\it (b)} images at 2.36, 4.86 and 8.46 GHz, u--v range = 3 -- 50 k$\lambda$,
$\theta_R = 6^{\prime\prime} \times 3^{\prime\prime}$, p.a. 0$^{\circ}$;

\noindent
{\it (c)} images at 4.86 and 8.46 GHz, u--v = range 6 -- 250 k$\lambda$
$\theta_R = 1^{\prime\prime} \times 0.5^{\prime\prime}$, p.a. 1$^{\circ}$.

%
\begin{table*}
\caption{Image details}
\label{data}
\begin{tabular}{lllccccc}
\hline
 &&&&&&&\\
 Source & RA$_{J2000}$ & DEC$_{J2000}$ & $\nu$ & FWHM, P.A. & rms & 
S$_{peak}$ & S$_{tot}$ \\
\\
 & h,$~$m,$~$s & $^{\circ}~~^{\prime}~~^{\prime\prime}$ & GHz 
& arcsec,$~~^{\circ}$ 
&  mJy/beam & mJy/beam & mJy \\
 &&&&&&&\\
\hline
 &&&&&&&\\
Halo        & 13 33 32.0 & $-31$ 41 00  & 1.40 & 41.9$\times$35.1, 55 &
  0.05 & -- & 20.0$^{\star}$ \\
J1333--3141 & 13 33 31.6 & $-$31 41 02  & 0.33 & 27.0$\times$12.5, --52 &
  4.04 & 98.9 & 259.0 \\ 
J1333--3141 & 13 33 31.6 & $-$31 41 02 & 4.86 & 1.46$\times$0.84, --57 &
0.05 & 1.25 & 25.1 \\
J1333--3141 & 13 33 31.6 & $-$31 41 02 & 8.46 & 0.88$\times$0.51, --53 &
0.04 & 0.64 & 15.6 \\
 &&&&&&&\\
\hline
 &&&&&&&\\
\end{tabular}
\\
$^{\star}$ flux density of the radio halo after subtraction of the point
sources and of the head--tail radio galaxy J1333--3141.\\
\end{table*}

\subsection{Morphology and integrated spectrum}

The high resolution images at 4.86 and 8.46 GHz (Figs. \ref{j1333c} and 
\ref{j1333x}) suggest that the observed morphology of J1333--3141 
is the result of projection effects coupled with strong interaction 
with the intracluster gas.

The northernmost and most compact knot, visible at both frequencies, 
is coincident
with the nucleus of the host galaxy and has the flattest
spectrum (see  Section 2.3). Therefore we conclude that it hosts 
the radio core. In Fig. \ref{j1333x} the nucleus, labelled as
component C, clearly shows up because of its flat spectrum and to the 
higher resolution, and the continuous bend of the northern jet is 
evident. The northern jet bends by 180$^{\circ}$ within the first 
kpc from the core, and its tail blends with the southern one.
The radio tail bends smoothly eastwards at
$\sim~25^{\prime\prime} - 30^{\prime\prime}$ from the core 
($\sim 32~-~38$ kpc), 
as first noted in the ATCA 1.38 and 2.36 GHz images (V2000).

We note the similarity between J1333--3141 and the head--tail
radio galaxy associated with NGC\,4869 at the centre of the
Coma cluster, whose jets bend sharply well within the optical
counterpart (Feretti et al. 1990).

\begin{figure} [!ht]
\resizebox{\hsize}{!}{\includegraphics{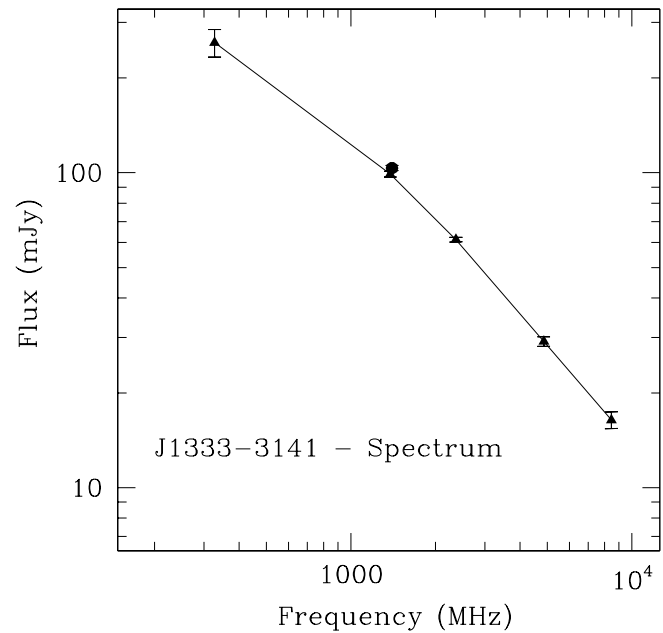}}
\caption{ Radio spectrum of J1333--3141. The 1.38 GHz and 2.36 GHz
flux density values are taken from V2000. Filled circle is the NVSS value
at 1.4 GHz.} 
\label{j1333spix}
\end{figure}

Using the data presented in this paper (see Section 2.1) and those 
published in V2000, we derived the total spectrum of the source in the 
range 330 MHz -- 8.46 GHz (see Fig. \ref{j1333spix}). 
We note the good consistency
between the full resolution 1.38 GHz and the NVSS 1.40 GHz 
flux density measurements, despite the difference in resolution.

%
%
\begin{table}
\begin{center}
\caption{Summary of the images used in Section 2.3}
\begin{tabular}{c|c|c|c|c|c}
\hline
 &&&&&\\
 u--v range & 330  & 1.38 & 2.36 & 4.86 & 8.46 \\ 
k$\lambda$  & MHz  & GHz  & GHz  & GHz  & GHz \\
 &&&&&\\
\hline
 &&&&&\\
0.5 -- 15 & X & X & X &   &    \\
3 -- 50   &   &   & X & X & X  \\
6 -- 250  &   &   &   & X & X  \\
 &&&&&\\
\hline
\end{tabular}
\end{center}
\end{table}

The spectrum of J1333--3141 can be considered a power law for 
$\nu > 1.38$ GHz, with $\alpha_{1.38}^{8.46}~=~0.99\pm0.06$, and it 
flattens to $\alpha_{0.33}^{1.38}~=~0.67\pm0.07$.

In order to estimate the flux density of the head--tail at 843 MHz
we fitted its total spectrum with a continuum injection model, and 
obtained S$_{843~MHz} = 146 \pm 6$ mJy.


\subsection{Spectral evolution along the tail}

We studied the evolution of the spectrum along the tail of J1333--3141
using the proper set of images (see Section 2.1 and Table 3) and 
the software Synage (Murgia 2001). The analysis allowed us to identify the
nucleus of the radio emission with the northernmost knot, coincident
with the central part of the associated galaxy (Fig. \ref{j1333c}). 
The spectral index in the radio core (see Fig. \ref{j1333x}), 
is $\alpha_{4.86}^{8.46} = 0.44 \pm 0.07$.

The most important signature in the synchrotron spectrum along the tail
is the presence of a frequency break.
We could place only a lower limit to the break frequency within
the first $\sim$ 8 kpc from the core, i.e. $\nu_{br}~>10~$GHz.
We considered three more regions along the ridgeline of the jet, 
chosen to be independent and to have good signal--to--noise ratio 
on all the images, and found that the break frequency progressively 
lowers at increasing distance from the nuclear region. 
Our results are reported in Table 4 and can be 
interpreted as the combination of electron aging and magnetic field 
decrement due to the jet expansion.

We note that the spectra
were fitted with the model discussed in Jaffe \& Perola (1974), thus
assuming that there is a redistribution of the electron pitch angles
on timescales short compared to their radiative life.
In order to avoid the uncertainties in the source volume evaluation
and the unknown evolution of the magnetic field with time,
we computed the electron radiative age t$_{rad}$ along the tail 
(reported also in Table 4) considering a constant magnetic field,
computed over the whole source volume (B$_{eq} = 5 \mu$G). 
We note that if we took into account the magnetic field decrement 
in the calculation of the electron radiative age, the values
given in Table 4 would be overestimated by up to a factor of 3, 
depending on how the magnetic field varies along the tail.

%
\begin{table}
\begin{center}
\caption{Radiative ages along the tail of J1333--3141}
\label{data}
\begin{tabular}{cccc}
\hline   
 &&&\\
 Dist1 & Dist2 & $\nu_{br}$ & t$_{rad}$ \\ 
\\
 arcsec & kpc  &   GHz      & $10^7$ yrs \\
 &&&\\
\hline
 &&&\\
8  & $\sim$ 10 & $>~10$ & $<~3$ \\
18 & $\sim$ 23 &  7.4 & 3.4 \\
23 & $\sim$ 30 &  4.4 & 4.5 \\
50 & $\sim$ 64 &  2.4 & 6.0 \\
 &&&\\
\hline
 &&&\\
\end{tabular}
\end{center}
Note to Table 4. \\
In the nuclear region (core and inner jets) the spectrum
is a powerlaw with $\alpha_{2.38}^{8.46} = 0.54 \pm 0.06$.\\
$\nu_b$ was computed assuming a magnetic field B = 5 $\mu$G
(see Section 2.3).
\end{table}

\section{The radio halo}

\subsection{Radio Observations}

We observed A3562 at 1.4 GHz with the VLA in the DnC configuration on 
10 July 2000, as part of a larger project whose aim is the study 
of the merging region between the centres of A3558 and A3562 at 
sub--mJy flux density levels.
In order to obtain uniform sensitivity, the observations were carried out 
in mosaic mode, covering the region of interest with four fields.
The total length of the observations was 4 hours, and we 
switched among  the four fields every 4.5 minutes. 
In Table 1 we report the coordinates and observing parameters 
only for the field relevant for the present paper.

The data reduction was carried out with the NRAO Astronomical
Imaging Package System (AIPS).
Each field was calibrated and imaged separately, then the four
images were combined together with the task LTESS. The rms in the 
final mosaic image is  $\sim 60 \mu$Jy/beam, and
the resolution is FWHM $\sim 41.9^{\prime\prime} \times 35.1^{\prime\prime}$,
in p.a. $55^{\circ}$.

Further details concerning the observations and data reduction will be
presented in Venturi et al. (2003).

\subsection{The morphology}

\begin{figure} [!ht]
\resizebox{\hsize}{!}{\includegraphics{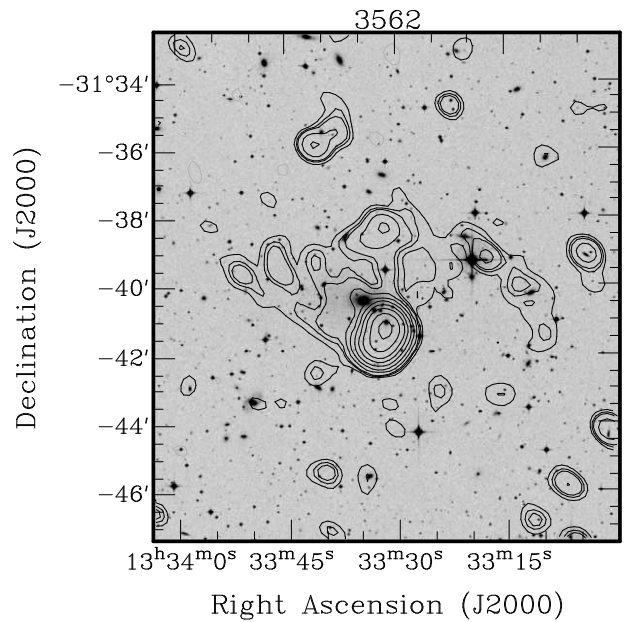}}
\caption{ 1.4 GHz radio image of the central radio halo in A3562
overlaid on the optical red image taken from the DSS--2. The FWHM
of the restoring beam is $41.9^{\prime\prime} \times 35.1^{\prime\prime}$,
in p.a. $55^{\circ}$. The radio contours are: $-$0.175, 0.175, 0.3,
0.4, 0.8, 1.6, 3.2, 6.4, 12, 25, 50 mJy/beam. }
\label{radiott}
\end{figure}

The portion of our final mosaic image including the halo 
emission at the centre of A3562 is shown in Fig. \ref{radiott}, overlaid 
on the optical red frame taken from the Digitised Sky Survey DSS--2. 
Details of the image are given in Table 2.

\begin{figure*} [!ht]
\resizebox{\hsize}{!}{\includegraphics{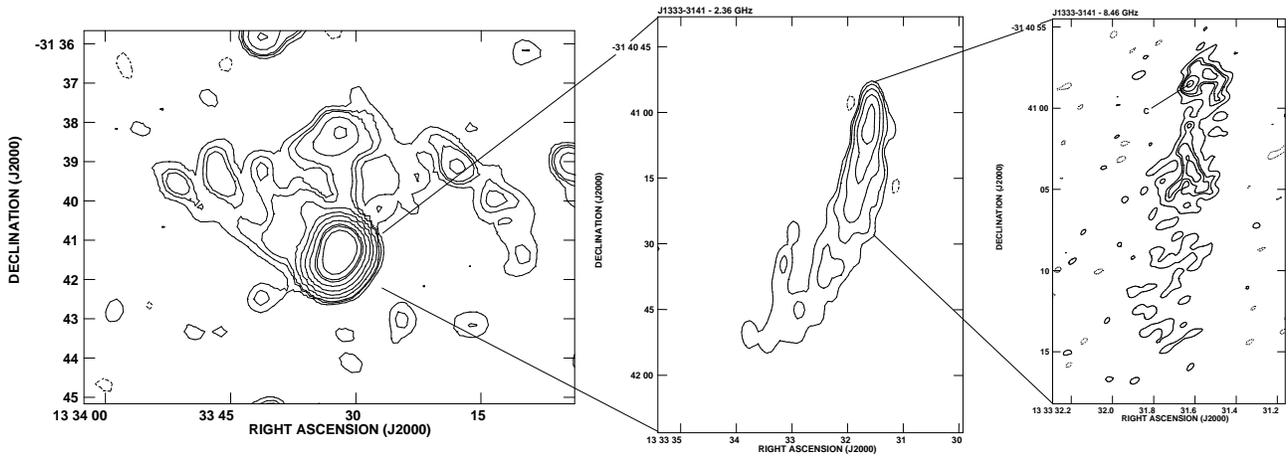}}
\caption{ Composite image of the radio halo and of J1333--3141.
Left and right panel are respectively the same as Figs. \ref{radiott} and
\ref{j1333x}. Middle panel: ATCA 2.36 GHz image of J1333--3141 (V2000).
Contour levels are 0.75 $\times$ ($\pm$0.6, 1.2, 2.5, 5, 10) mJy/beam,
and the restoring beam is $5.43^{\prime\prime} \times 3.16^{\prime\prime}$,
p.a.0.7$^{\circ}$.} 
\label{halo_tail}
\end{figure*}

\begin{figure} [!ht]
\resizebox{\hsize}{!}{\includegraphics{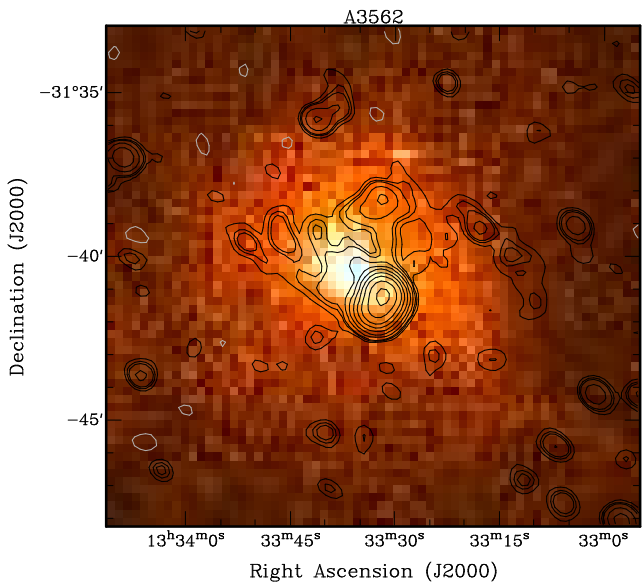}}
\caption{ 1.4 GHz radio image of the central radio halo in A3562
overlaid on the ROSAT All Sky Survey image. Details of the
radio image are the same as in Fig. \ref{radiott}.}
\label{radiox}
\end{figure}

The radio halo at the centre of A3562 has low surface brightness
($\le~0.175$ mJy/beam, i.e. $\sim 7\times10^{-5}$ mJy/arcsec$^2$)
and irregular shape, and encompasses the 
head--tail radio galaxy J1333--3141, as clear from 
Fig. \ref{halo_tail}. 
The overall morphology 
recorded in our images is in very good agreement with the
NVSS image. The largest angular size is $\sim~8^{\prime}$, corresponding
to $\sim$ 620 kpc. 
The extension of the radio halo is smaller than that of the 
X--ray emission, imaged by ROSAT, however it is coincident with 
its brightest part, as clearly visible from Fig. \ref{radiox}.
Together with those found in A1300 and A2218
(Feretti 2000), it is one of the smallest and faintest
radio halos known to date.

We note that the observed size of halos depends both on
the sensitivity of the observations and on the intrinsic brightness 
distribution. As a consequence, it is possible that we are 
underestimating 
the size of our halo, as the presence of positive residuals
south of the halo and of J1333--3141 may suggest 
(see Fig. \ref{radiott}). A tool to estimate the intrinsic size of 
the radio halo in A3562 is the radio and X--brightness relation
(Govoni et al. 2001b). However, our attempt in this direction 
was unfruitful since this relation shows broad dispersion at low 
brightness levels, and its slope is not unique.

In order to estimate the flux density of the halo, we 
subtracted the point--like components from the u--v data, and obtained 
a value S$_{1.4~GHz}~=~20~\pm$ 2 mJy. Given the source extension, 
the flux density error is dominated by the thermal noise 
rather than by the calibration uncertainties.  
The radio power is P$_{1.4~GHz} = 2~\times~10^{23}$ W Hz$^{-1}$. 

We note that the western part of the halo has a filamentary
shape, which might be suggestive of a faint relic. It contributes
with $\sim 1.8$ mJy to the total flux density of the halo. 
However, from the 
present radio observations and from the X--ray information available 
in the literature (Ettori et al. 2000), it is impossible to test if
the imaged radio emission is the result of 
a radio halo and a relic. Awarded high resolution Chandra X--ray 
observations and low frequency radio observations will provide
the critical information necessary
to interpret the observed radio features.

\subsection{The radio spectrum}

We estimated the spectral index of the radio halo 
in the frequency range 843 MHz -- 1.4 GHz using its flux density 
at 1.4 GHz reported in Table 2, the radio spectrum of J1333--3141
and the MOST 843 MHz image shown in V2000. After subtracting the
contribution of J1333--3141 to the 843 MHz image, we obtain
S$_{843~MHz}~=~59 \pm 6$ mJy.
We note that the extension of the radio halo at 1.4 GHz 
(Fig. \ref{radiott}) is larger than at 843 MHz, due to its low surface 
brightness and to the better sensitivity of the 
1.4 GHz data. Considering all the uncertainties, we conclude that 
$\alpha_{843~ MHz}^{1.4~GHz}$ is in the range $\sim$ 1.9 -- 2.3.

\section{Discussion}

\subsection{The radio halo and the cluster properties}

The total power and linear size of the halo are respectively 
P$_{1.4 GHz} = 2 \times 10^{23}$ W/Hz and $\sim$ 620 kpc. These
values are smaller than those found for the other halos known in the 
literature. For comparison, the cluster radio core and the 
cooling flow region are both of the order of 1.2 arcmin, i.e. 
$\sim 90$ kpc (Peres et al. 1998; Ettori et al. 2000).

Assuming that standard equipartition conditions hold in the source 
($\Phi=1$, k=1, $\nu_1$ = 10 MHz, $\nu_2$=100 GHz), the magnetic field 
in the halo is B$_{min} \sim 0.4~ \mu$G, and the internal energy
u$_{min} \sim 1.5 \times 10^{-14}$ erg cm$^{-3}$. These values
are in the range found for the halos known to date (Feretti 2000).

\begin{figure*} [!ht]
\resizebox{\hsize}{!}{\includegraphics{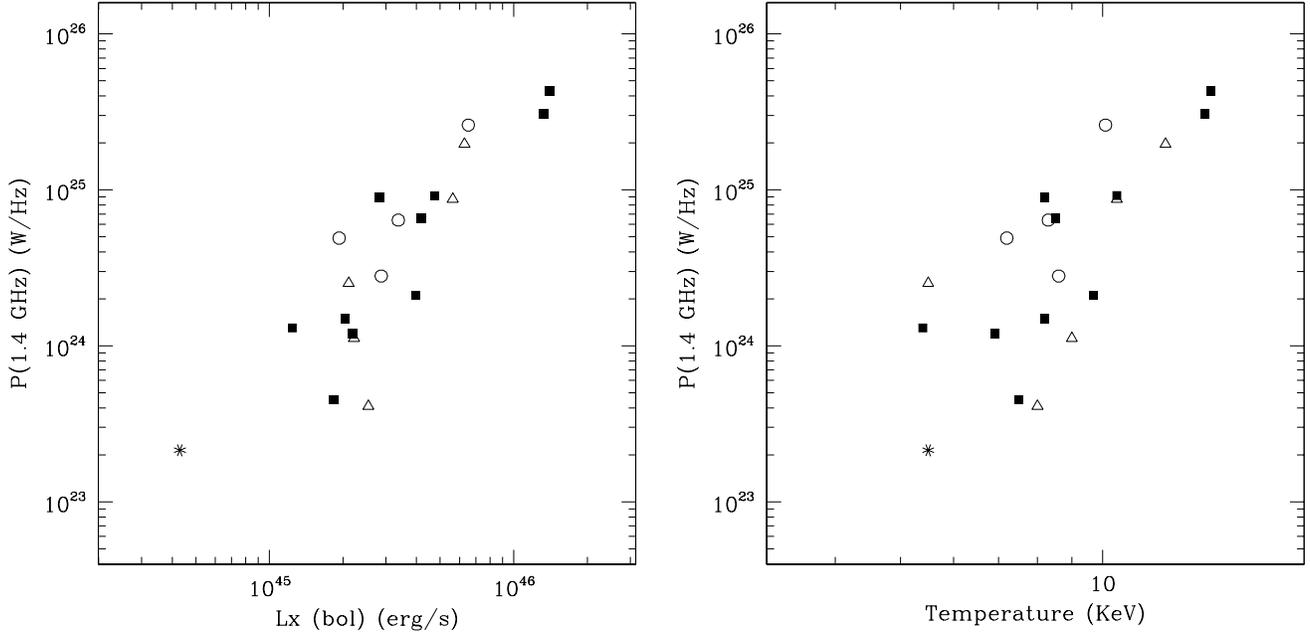}}
\caption{ logL$_X$ -- logP$_{1.4 GHz}$ diagram (left) and 
Temperature -- logP$_{1.4 GHz}$ diagram (right) for the cluster
radio halos known to date, regardless of their size. 
Symbols are as follows: 
filled squares from Feretti (2000); open circles from Govoni et
al. (2001a); open triangles from Bacchi et al. (2003); star is A3562.}
\label{correl}
\end{figure*}

In Fig. \ref{correl}, left and right panel respectively, 
we show the 
logL$_X$ (bolometric) -- logP$_{1.4 GHz}$  relation and the 
Temperature -- logP$_{1.4 GHz}$  relation for the halos in the 
literature (note that the plot shows only halos, i.e.
relics and mini--halos were not included) 
and for A3562, whose X--ray data were derived from
Beppo-SAX observations (see Section 1, Ettori et al. 2000).
The figure indicates that the halo in A3562
fills a new region of the two diagrams, and nicely fits the
extension of the correlation found by Bacchi et al. (2003) for
large size halos. In particular, we note that the inclusion
of A3562 in their correlation is consistent, within the errors,
with their fit.
This result suggests that 
radio halos exist and can be detected also in clusters with 
L$_X~<~10^{45}$ erg s$^{-1}$,
and that the correlation linking the
X--ray properties and the radio power holds over three orders of
magnitude in logP$_{1.4 GHz}$ and one order of magnitude in temperature
and X--ray bolometric luminosity.
We note that many faint radio halos with surface brightness lower
than in A3562 are expected according to the 
the radio halo luminosity function study of En{\ss}lin \& R\"ottgering
(2002), which was based on an extrapolation of the observed
L$_X$  -- P$_{1.4 GHz}$ relation. Our measurements seem to
support the validity of this extrapolation.

%
%

\subsection{Origin of the halo}

The radio power and size of the halo in A3562 lead to a fundamental
question: has the source already fully developed or is it 
still growing at the present time?

\begin{figure} [!ht]
\resizebox{\hsize}{!}{\includegraphics{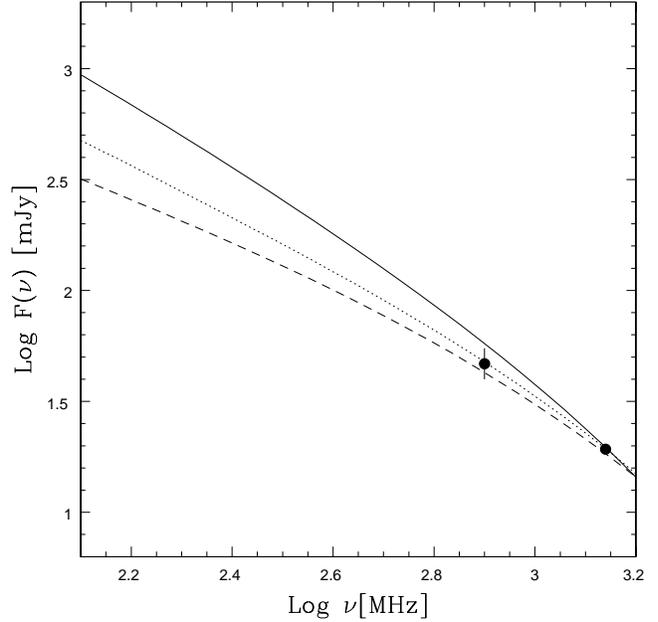}}
\caption{ Synchrotron halo spectra from a reaccelerated
electron population are superimposed on the radio data (filled
circles) for A3562 at 1.4 GHz and 843 MHz (this paper).
The models are performed (assuming $B < 3 \mu$G) to match the 1.4 GHz
flux and to reproduce the observed halo radial extension.
The solid line corresponds to a reacceleration time of 0.1 Gyr, 
the dashed line to 0.4 Gyr and the dotted line to an intermediate value.}
\label{halo_spect}
\end{figure}

The region of the logP -- logL$_X$ correlation where A3562 is located, 
is of crucial importance 
for testing current models of the halo formation, in particular the 
two--phase model (Brunetti et al. 2001), which includes an 
initial injection of relativistic particles (phase 1), and subsequent 
reacceleration due to turbulence powered by cluster merger (phase 2). 
In the framework of this model, A3562 may be interpreted either as
{\it (a)} the result of a modest reacceleration phase, or as 
{\it (b)} a radio halo at the beginning of an energetic reacceleration 
process. 
In both cases, A3562--like halos are crucial to our understanding of 
the merger--halo connection.
In particular, radio halos in the logP -- logL$_X$ region 
occupied by A3562 are predicted to have a spectrum in the range 
1.5 -- 2 at frequencies lower than 1.4 GHz. 
The spectral index of A3562 $\alpha_{843~MHz}^{1.4~GHz}$ in the 
range $\sim$ 1.9 -- 2.3 (Section 3.3). However this frequency range is not 
wide enough to discriminate
between the two possibilities {\it (a)} and {\it (b)}. In Fig. \ref{halo_spect}
synchrotron radio spectra from a reaccelerated electron population are
shown. The plot shows that the spectra flatten with increasing duration of the 
reacceleration phase, going from 0.1 Gyr (solid line) to 0.4 Gyr 
(dashed line). For a given duration of the reacceleration
phase a qualitatively similar trend is also obtained increasing the 
acceleration efficiency. 
From Fig. \ref{halo_spect} it is clear that additional
observations at very low frequency (few hundreds of MHz)
will be of crucial importance to confirm this.

\subsection{Electron injection in the halo}

The head--tail radio galaxy J1333--3141 is completely embedded
in the halo emission (see Fig. \ref{halo_tail}), and it is the best 
candidate for the replenishment of the electron population in the 
central part of the cluster. 
We computed the number of electrons N$_{e^-}$ the radio galaxy 
injected in the intracluster medium in order to see if it is high 
enough to feed the halo.
We carried out our calculation considering the formulae
for the equipartition magnetic field given in Brunetti (2002),
where low energy electrons are also taken into account 
(B$_{eq}$(J1333--3141) $ \sim 8.0 \times 10^{-6}$ G). 

\noindent
We derived 
N$_{e^-}$ (J1333--3141) 
$\sim 9 \times 10^{59} ({{\gamma_{min}} \over {10}})^{-1.46}$,
where $\gamma_{min}$ is the minimum Lorentz factor of the radiating electrons.

Based on the two--phase models used to reproduce the spectrum shown in 
Fig. \ref{halo_spect} we estimated the minimum number of relic electrons 
necessary to obtain the observed radio halo, i.e. 
N$_{e^-}({\rm halo}) \sim 6 \times 10^{60}$. 
Assuming that the tail is the only source of electrons injected in the 
halo, it should be 
N$_{e^-}$(halo) $\sim$ t$_{min} \times$  
N$_{e^-}$(J1333--3141)/ t$_{dyn}$, where t$_{dyn}$ is the time 
the galaxy needed to cross the length of the radio tail. We thus 
derive that the minimum injection time t$_{min}$ is:

\begin{equation}
{\rm t}_{min} \sim 3 \times 10^8 \left({{\gamma_{min}} \over {10}}\right)^{-\chi} 
\left({{\rm v} \over {1000~~{\rm km~s^{-1}}}}\right)^{-1}
\Delta^{-(3+\alpha)}~~~~{\rm yr}
\end{equation}

\noindent
where $\chi=2(\alpha^2+5\alpha-1)/(3+\alpha)$,
$\Delta = {\rm B}_{eq}/{\rm B}$ and v is the velocity of the tail 
(km s$^{-1}$).

This value of t$_{min}$ should be compared with the time required
for the host galaxy to cross the radio halo region ($\sim$ 600 kpc), 
i.e. t$_{cross} \sim 6 \times 10^8 ({\rm v}/1000)^{-1}$ yrs.
If we assume that v is of the order of the dispersion velocity
in the cluster ($\sigma_{\rm v}$ = 987 km s$^{-1}$, Bardelli et al. 1998),
equipartition conditions ($\Delta$ = 1), $\alpha=0.65$, and a typical
value for $\gamma_{min} \sim 10 - 30$, we conclude that the tail must be
active for a considerable fraction of the crossing time.

If the above scenario is correct, the reacceleration phase started 
less than few $\times 10^8$ yrs ago, leading us to the conclusion
that this radio halo is young and at the beginning of a reacceleration.

\section{Conclusions}

In this paper we present a thorough multifrequency study of the radio
halo, located at the centre of the merging cluster of galaxies A3562,
and of the head--tail radio galaxy J1333--3141. 
Our results can be summarised as follows:

\noindent
-- the halo has a steep spectrum, with $\alpha_{843~MHz}^{1.4~GHz} \sim 2$,
an equipartition magnetic field B$_{min} \sim 0.4~\mu$G and internal energy
u$_{min} \sim 1.5 \times 10^{-14}$ erg cm$^{-3}$;

\noindent
-- its size and radio power at 1.4 GHz are amongst the lowest
found for this class of sources and they nicely fit the correlations
with other cluster parameters (such as bolometric X--ray luminosity
and temperature) found in the literature, extending them to
lower values for all quantities involved (Feretti 2000; Govoni et al. 2001a);

\noindent 
-- the total number of electrons injected in the central region of
the cluster by the head--tail radio galaxy J1333--3141 are enough
to feed the radio halo if we assume that the galaxy has been radio
active over a considerable fraction of its crossing time 
(t$_{cross} \sim 6\times10^8$ yrs);

\noindent
-- in the framework of the reacceleration models, the scenario in A3562 
could be due either to a modest reacceleration phase, or to an initial 
stage of an energetic reacceleration process. The connection between the
radio halo and J1333--3141 might suggest that this second possibility
is best matched to the observations. However very low
frequency radio observations are necessary to discriminate between the
two possibilities.

\begin{acknowledgements}
The authors wish to thank the referee, Dr. T. En{\ss}lin, for fruitful
comments. Thanks are due to Dr. M. Murgia for insightful discussion and
help in running the program Synage.
G.B. acknowledges financial support by CNR, under grant CNRG00CF0A.
This work has been partially supported by the Italian Space Agency grants
ASI-I-R-105-00 and ASI-I-R-037-01, and by the Italian Ministery (MIUR)
grant COFIN2001 ``Clusters and groups of galaxies: the interplay between
dark and baryonic matter".
NRAO is a facility of the National Science Foundation, operated under
cooperative agreement by Associated Universities, Inc.
This work has made use of the NASA/IPAC Extragalactic Database NED
which is operated by the JPL, California Institute of Technology, 
under contract with the National Aeronautics and Space Administration.
\end{acknowledgements}


\end{document}